# Analytics-Driven Digital Platform for Regional Growth and Development: A Case Study from Norway

Salah Uddin Ahmed, Steinar Aasnass, Fisnik Dalipi, and Knut Hesten
University of South-Eastern Norway
Hønefoss, Norway
e-mail: {salah.ahmed, steinar.aasnass, fisnik.dalipi}@usn.no,
knut.hesten@gmail.com

*Abstract*—**In this paper, we present the growth barometer (Vekstbarometer in Norwegian), which is a digital platform that provides the development trends in the regional context in a visual and user-friendly way. The platform is developed to use open data from different sources that is presented mainly in five main groups: goals, premises or prerequisites for growth, industries, growth, and expectations. Furthermore, it also helps to improve decision-making and transparency, as well as provide new knowledge for research and society. The platform uses sensitive and non-sensitive open data. In contrast to other similar digital platforms from Norway, where the data is presented as raw data or with basic level of presentations, our platform is advantageous since it provides a range of options for visualization that makes the statistics more comprehensive.**

## I. Introduction

Nowadays, open data plays an indispensable role for governments' strategy to deal with many innovation challenges of the future. Furthermore, open innovation philosophies and approaches are being launched and adopted by public sectors in many different countries [1][2]. In Norway, many governmental agencies have embraced the open data initiative and are making data available for public use. Hence, businesses and citizens can now access and utilize these open data resources to create innovative value-added products and services [3]. When it comes to defining the meaning of open data, we use the definition provided by Open Data Institute, which defines open data as "data that is made available by government, business and individuals for anyone to access, use and share" [4]. The global economic potential value of open data has been estimated to $3 trillion [5]. On the other hand, the potential and advantage of Open Government Data (OGD) for enhancing services in different economic sectors has not been realized to a large extent [6].

In this paper, by using open data provided by *Statistics Norway* [7], *Real Estate Norway* [8] and *the Brønnøysund Register Centre* [9], we present the *Vekstbarometer* digital platform, an analytics-driven web application, which contains regional indicators presented along with research-based knowledge relevant to regions' growth. According to the conceptual model presented in Figure 1, regional growth is associated with growth in Value Creation, Employment, Workplace and Population. A region's growth is often measured by growth in GDP (Gross Domestic Product_. However, the aim of the policy is to contribute to higher welfare and transform the region into better place to live and run business. In order for the Growth Barometer to be able to explain developments and provide relevant information related to political and business decisions, a broader definition of growth is therefore needed. The indicators are based on open data while statistical visualizations can be generated also for purposes other than regional growth. We have focused our study in the region of Ringerike from the southeastern part of Norway, which has recently seen a decline in the number of jobs and weak economic growth. To the best of our knowledge, there is no similar open-data based system previously developed or introduced, which encompasses research-based knowledge on a regional level in Norway. Nationally, there exist different digital platforms, but they are all different because not only they are using different data categories, but they also have different purposes. Several regions have business barometers based on survey data, collected on annual basis and register data. They forecast the national and regional business trends. One example is Konjukturbarometer Østlandet [10], a digital platform that contains, among others, a knowledge-based database on developments in the counties of Hedmark, Oppland, Oslo, and Akershus. The Confederation of Norwegian Enterprise (NHO), which is Norway's largest organization for employers and the leading business lobbyist, is another example of digital platform. Their platform, Økonomibarometeret [11] covers the market situation, operating profit, investments, and employment on a county and national level.

The main intention behind developing our innovative system, which can be accessed at *vekstbarometer.usn.no,* is to provide public sector authorities and local industries a management tool with key indicators related to the region's growth. Apart from having the potential to enhance public's sector transparency and engagement of civil society, our system can also contribute towards improving economic growth through processing and illustrating regional open data in a comprehensive way.

The rest of paper is structured as following: Section 2 presents related work, Section 3 highlights the need of digital platform for regional development in Ringerike, while Section 4 introduces the *Vekstbarometer* system and the technology used for its development, along with its strength and impact for regional growth. Lastly, Section 5 concludes the paper and provides some insights about potential future work.

## II. RELATED WORK

Due to the emergence and pervasiveness of ICT (Information and communication Technologies), many governments across the globe have been undertaking initiatives to transform themselves into e-governments [12], and subsequently are encouraging citizen participation in governments. OGD is one of the main extensions of such e-government initiatives [13]. OGD is making data freely available and accessible to all with the goal to ensure public accountability and transparency [14], to empower innovation in different economic sectors and to enhance efficiency in administration.

Nevertheless, in order for stakeholders to derive the public value out of the open data, it is of paramount importance for the data sets to be re-usable, comprehensive, interpretable, complete, and permit user-friendly interface. Moreover, government authorities should be proactive towards ensuring that the data sets are published according to stipulated norms, such as protecting personal and private information of the users, or prohibiting the release of sensitive data related to national security. In this perspective, the Norwegian government has created a license for open government data and have recommended all data owners in the public sector to apply the license, which contains, inter alia, information on preserving confidential and personal data [15]. When it comes to global status and trends of open data in the context of readiness, implementation and impact, Norway ranks among the top ten best countries in the world [16].

In the literature, there are two categories of OGD research. The first category is mainly based on conceptual frameworks and theoretical proposals [17][18], also including studies discussing the main stages of the OGD life-cycle [19]. The second category, where our work belongs, includes studies that are conducted in different countries by using OGD at the national or local level. Furthermore, this category is characterized by open data exploration and exploitation, where data from multiple sources of government agencies are processed and visualized for further use, such as to conduct various analysis, create mashups, to enhance the interpretability of open data, or even innovate upon the open data.

Over the past years, multiple applications have been developed based upon the open data sets across the world, focusing primarily on larger cities, such as Chicago [20][21], New York [22], Dublin [23], St. Petersburg [24], Singapore [25].

In spite of the interest and the rapid proliferation of open data platforms, many challenges remain with the accessibility and usability of platforms using open data, data quality and completeness, and interpretability of open data. When it comes to enhancing the interpretability of open data, the authors of the research work [20][21] conduct a case study to analyze the open socioeconomic data released by the city of Chicago, where they apply different visualizations tailored for univariate, bivariate and multivariate analysis. This approach helped them to understand that exploring open urban data can lead to more effective data interpretation. Considering the usability for user interface design of open data platforms, a case study scenario involving a transportation challenge in Dublin City identified important patterns for highly usable open data platforms for open data policy, recommending these platforms should adopt user-friendly technology and social media platforms [23]. From architectural point of view, there is a work presented by [25]. Here, the authors developed an open data platform prototype and illustrate the requirements and the architecture of open real time digital platform to serve as a base for programming the city of Singapore, and provide visual data analytics in an urban context.

## III. THE NEED FOR A DIGITAL PLATFORM

Several key factors related to growth and welfare points to the wrong direction of development for the region Ringerike. The region has around 43,000 inhabitants while the number of total new employment for the region was only 145 people over the past 10 years [26]. The number of jobs decreased was 321 for the same period. At the same time, value creation for the region's business sector has shown a weaker development compared to other regions that can be naturally compared with Ringerike, i.e., neighboring regions with similar background and assets. If this negative trend is not reversed, the Ringerike community will face the consequences of declining private and public welfare. This will make the region a less good place to live. However, politicians and businesses in the region are optimistic in terms of the future. New high-speed railway from Oslo to Ringerike region, and the new four-lane highway from Oslo to Hønefoss is planned to be completed around 2028. This gives enormous opportunities for the region to reverse the negative trend and create new growth that will ensure future welfare and good living conditions in the region. Nevertheless, in order to achieve this growth, it requires good decisions from the region's public authorities and industry. A growth barometer that monitors the growth in the region and significant conditions for growth could provide a useful management tool for strategic decisions.

The target user group for the barometer will be politicians, municipal administration, business and the community. There are large amounts of statistics and information related to growth and development for municipality regions from different sources. However, there is a need to provide these statistics and information in a customized way targeted for the use in Ringerike region, develop new and better statistics, and break down the statistics at regional and local level. In the preparatory work, we have conducted a simple survey to find similar tools or platforms as growth barometer for other regions in Norway.

We have found a large number of solutions, but none with the approach and objectives that we mentioned here. In this context, we believe that a digital platform like growth barometer could give Ringerike region a significant competitive advantage over other regions that are also working towards regional growth and development.

### A. The logical model of Barometer

The conceptual model behind the growth barometer is given in the Figure 1 below.

## IV. VEKSTBAROMETER - THE DIGITAL PLATFORM TECHNOLOGY

The growth barometer (*Vekstbarometer* in Norwegian) is a digital platform that provides the development trends in the regional context in a visual and user-friendly way. The platform uses data from different sources that is presented mainly in five main groups: goals, premises or prerequisites for growth, industries, growth, and expectations.

Each group contains several categories and each category contains several variables, which actually contain the statistical data. The groups and the categories form the information architecture of the digital platform, which is shown in the Figure 2.

The group goal contains categories: Population, Value Creation, Employment, Jobs, and Welfare. Each of these

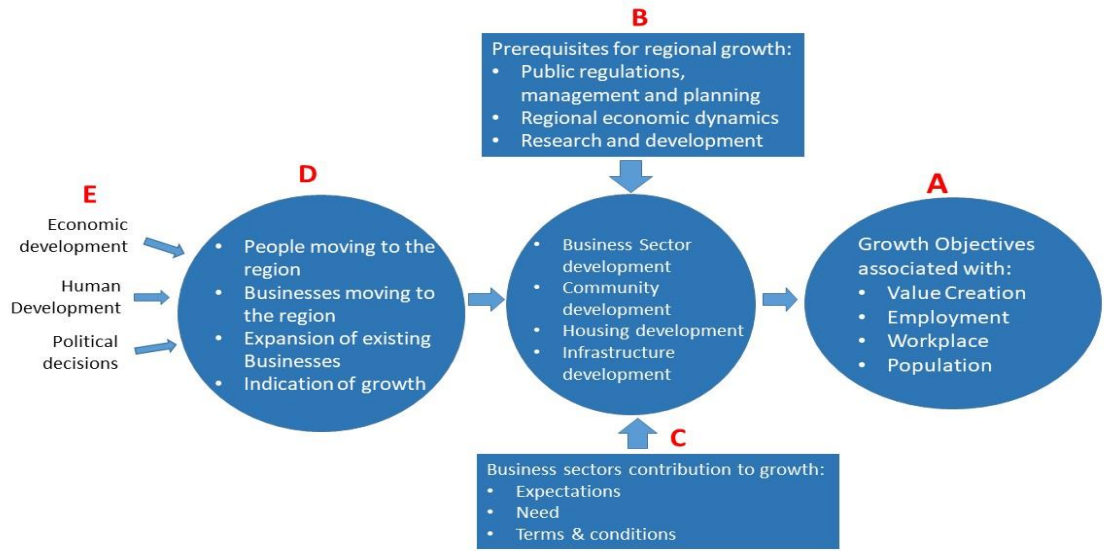

Figure 1. The logical model behind the vekstbarometer

The objective of regional development is given through the points in (A), i.e., higher value creation, employment, jobs and higher populations. Nationally, there will always be a certain amount of people and businesses considering to "relocate". There will be national competition to get these businesses and individuals to establish themselves in our region. Here, conditions for regional growth (B) and local industry's contribution to local growth (C) can serve as lucrative points for capturing a good share of the influx of national movements of persons and businesses. If the conditions (B) and the local business sector's efforts (C) are large, this portion will be large. The result will be a major influx of businesses and individuals. We assume that growth targets mentioned in (A) could be achieved together with regional economic dynamics linked to business development, housing construction, etc.

categories contain a number of variables that are not shown in the figure. These variables represent the data analyses and the visual representations in the form of charts, graphs, and diagrams. For example, the population category contains ten variables, some of which are: total inhabitants, age-wise population, population change trends, net population change etc. Each variable is represented by a number, for example total inhabitants (1), age-wise population (25). The numbers are not assigned in any order; rather they were assigned when the statistics of the variable were being added in the system.

The variable number can be seen from the URL; when browsing a certain statistics from the navigation menu the URL gets changed with the variable number.

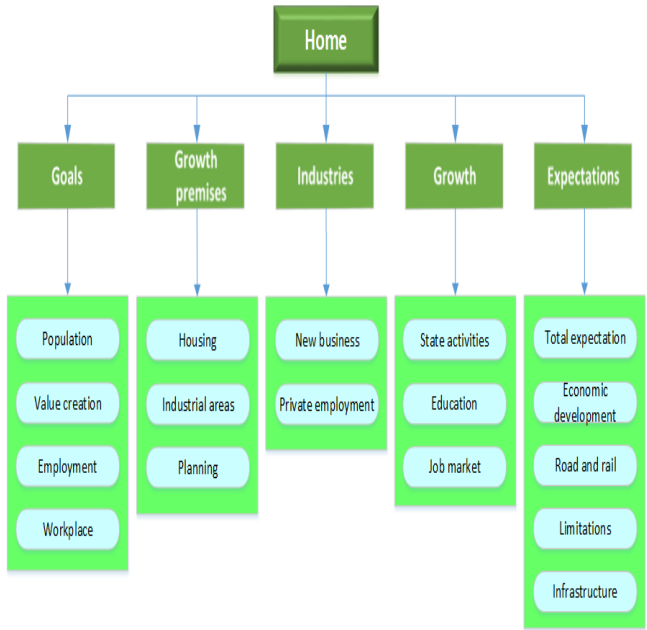

Figure 2. Information architecture of the application

Thus, the statistic variable 25 can be accessed from the navigation menu, as well as from the url by appending the variable number after the URL given at [27].

Apart from presenting the open data, *Vekstbarometer* also presents survey data from the local industries that reflects the expectations and assumptions of the local entrepreneurs and business owners. The application consists of mainly three parts: backend, presentation and database. The schematic architecture of the growth barometer application is given in Figure 3.

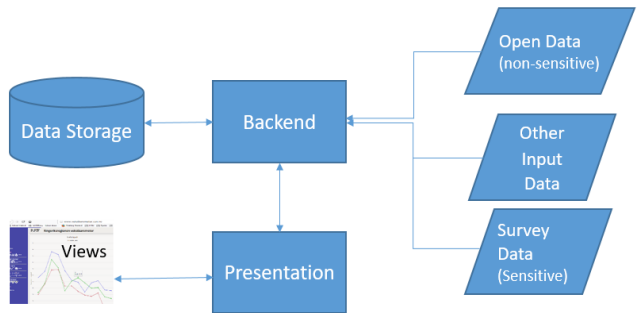

Figure 3. *Vekstbarometer* system architecture

Back-end's main feature is to process requests from the presentation layer. Based on the requests, it retrieves raw data from the database, sort the data and send them to the presentation. Backend retrieves data from external data storage such as open data that are provided by others through APIs (Application Programming Interface). It also processes other input data to the system and survey data. Data from external sources are retrieved in the desired formats and are saved in the database for storage. External data sources are fetched periodically to keep the main database updated. Presentation takes care of conversion of data transmitted from backend, transformation of these and display on-screen to end-user. This is the main part of the application that the user sees and interacts.

The database holds the persistent and transient data that are critical for running the application. We have used *MySQL* databases for the persistent storage and *Angular 2* for presentation and frontend, and *PHP* for Backend.

### A. Data

*Vekstbarometer* is a data visualization platform. It includes several kinds of data. Some data are sensitive and some are not sensitive. Data that are already presented as open data by other outlets, such as *Statistics Norway* are non-sensitive. Since these data are already available and open for public, we do not have any restrictions on showing them in raw format or in modified format. Apart from non-sensitive data, there are data or part of data that contains some sensitive information about people or businesses that should not be made publicly available, e.g. a person's personal number (social security number) or a business's sales and marketing plans, confidential customer or supplier information, etc. These data contain privileged or proprietary information that only certain people are allowed to see. In our system, we present data collected through surveys from the companies in the region, which contain sensitive information. However, these data are not represented in a way that are trackable to individual response level. Rather, the data in the system are presented only as aggregated data. We make sure that data does not contain any attributes that can directly identify a person's response. Identifiers, business names, or organization numbers are removed from the data set and the results are published only as aggregates.

Besides, we follow the GDPR (General Data Protection Rules) rules when collecting data from local businesses. Most of the data are collected from other sources by some operations such as filtering or making queries in order to fit our local needs that are represented by the statistics variables in our system. However, the barometer also shows some newly created data that are not available in any other data sources.

### B. Visualizations

Data is presented using different kinds of charts and diagrams. The diagrams include line charts, bar charts, stacked columns, stacked percentage column, column with drill down, pie charts etc. For visualizations, we have used a javascript library called *Highcharts*. With *Highcharts* is it is possible to create charts in many shapes, like heat maps, waterfall series and more. Additionally, the charts are highly configurable, customizable, and interactive.

By configurable and customizable, we mean that values can be added and removed on the charts on the fly. As an example, the Figure 4 below presenting variable 25, which shows age-wise population comparisons in different regions, we can choose the age groups that will be presented in the graph. The graph shows not only the percentage of an age group, but also the absolute value of that group in a tooltip text when mouse is hovered over that group. Besides the customization features, the regions in the x-axis are linked to a detailed view of the age wise population chart for only that region (Figure 5), thus making the visualization configurable and interactive at the same time.

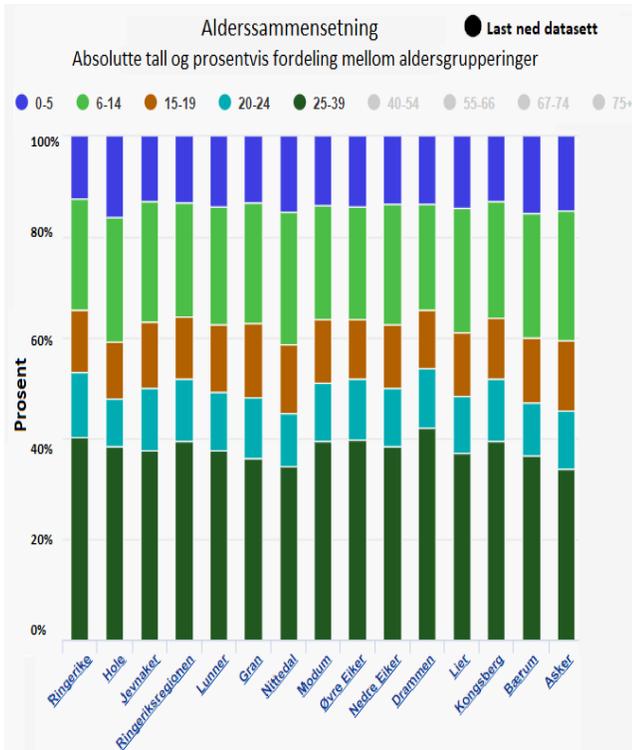

Figure 4. Age wise population chart for all regions, variable 25

The visualization of a statistic variable is further complemented by i) textual description of the variable, ii) links to related documents iii) links to related variables and iv) alternative graph options. The textual description gives a brief introduction about the variable for better understanding and use. Users can read the links to related documents for further concepts. By using the links to related variables, user can navigate to related statistic variable directly from this page.

Alternative graph option lists a number of options, and by selection one from those options, the user can view the same statistical variable in a different form of presentation, for example linear chart, bar chart, pie chart etc.

### C. The Strength and Impact of the platform

The digital platform shows data in an easy, comprehensible and meaningful way. Although some data already exist in other data outlets, our platform adds value by making a better presentation of the data.

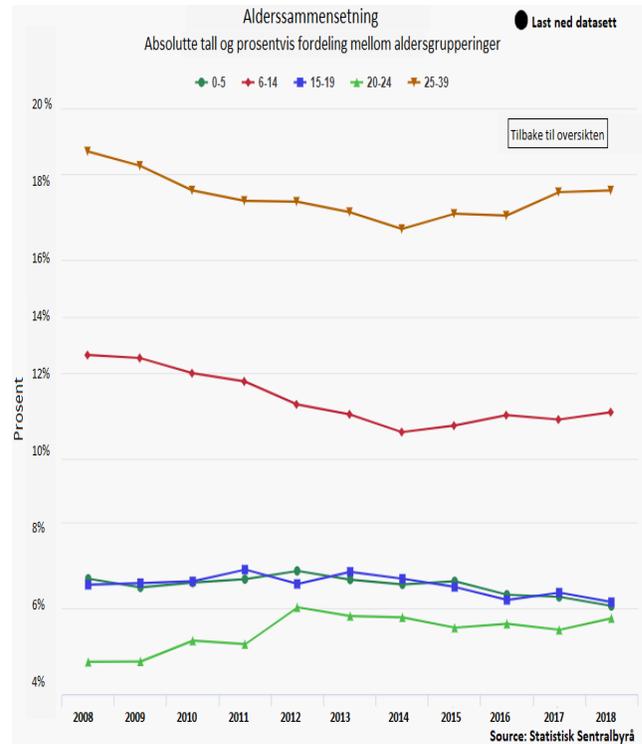

Figure 5. Age wise population chart for a specific region, variable 25

Furthermore, in contrast to other platforms, such as [28]-[32], where the data is presented as raw data or with basic level of presentations, our platform is advantageous since it provides a range of options for visualization that makes the statistics more comprehensive. Even in case where the platform is showing existing data from other sources, still we add value on it.

In addition to showing already existing data, the platform shows some new data. The new data can be of two types: i) newly formed data, ii) newly created data. Newly formed data is created by combining and filtering data from multiple sources where the data were partly available but the in the form that is presented here.

On the other hand, newly created data is the case where we create, collect or gather very new data that were not available or presented in any other platform. As an example of newly formed data we can mention "future population growth" which is represented in our platform as statistical variable 56 (see Figure 6). Here, the future expected population growth is taken from three different sources: *Statistics Norway* assumptions, political assumption and housing building assumption. This new variable shows the lack or surplus of housing capacities with political assumptions or the *Statistics Norway* assumption and gives a good indication if they are feasible or not compared to the housing capacities that are planned for the region. When it comes to the possibility of downloading data and images, besides showing the statistics in visual form, the platform also provides the option to download data in multiple formats, such as csv, pdf, xls, png and jpeg image, svg vector image, etc. It also provides the option of printing the chart and view the data of the chart in tabular forms. These options increase the usability of the tool and facilitates multiple use cases for the users of the system. Other users can use customized graphs and charts from our platform and include them in presentations or documents.

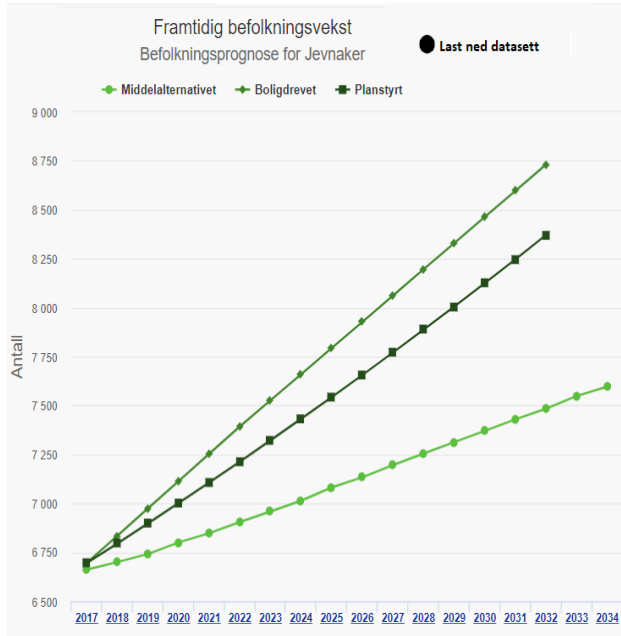

Figure 6. Future population growth, variable 56

The *vekstbarometer* platform is a constituent part of the five years long Growth Barometer project, which is managed by the Regional Development Group at the University of South-Eastern Norway (usn.no). Since the initiation of the project, a status report is issued continuously presenting the current regional growth. The Ringerike region also has a business policy strategy that sets the premises for how business policy is to be pursued to create growth in the region's business community. The strategy is created based on *Vekstbarometer* platform and presents the annual results of the regional growth according to the growth objectives given at Figure 1. Moreover, the municipality authorities can rely on vekstbarometer data in order to define their priorities and make better decisions.

## V. Conclusion and Future Work

In this paper, inspired by the recent developments in the field of open data initiatives, we have presented the *Vekstbarometer* digital platform. We developed the platform to combine multiple open data sources to generate various visualizations. This gives insights into the regional growth and development, and demonstrates the usefulness of open data in regional context. Furthermore, it also helps to improve decision-making and transparency, as well as provide new knowledge for research and society. The platform uses sensitive and non-sensitive open data.

As a future work, we intend to expand the functionality of the platform by expanding the dataset and focus on converting the digital platform into a fully-fledged and mobile-friendly application.

## Acknowledgment

The authors greatly acknowledge the grant from Sparebank1 Ringerike Hadeland, which kindly supported this project, and the important contribution given by Emil Kalstø, who performed some programming tasks.